\documentclass[showpacs,amsmath,amssymb,prl,superscriptaddress,footinbib,twocolumn]{revtex4-1}
\usepackage{amssymb}
\usepackage{graphics}
\usepackage{epstopdf}
\usepackage{epsfig}
\usepackage{dcolumn}
\usepackage{bm}
\renewcommand{\vec}[1]{\mathbf{#1}}
 
\newcommand{\abs}[1]{\left| #1 \right|} 
 
\let\baraccent=\= 
\renewcommand{\=}[1]{\stackrel{#1}{=}} 

\usepackage{color}
 \definecolor{blue}{rgb}{0,0,1} 
 \definecolor{sepia}{rgb}{0,0.8,0.2}
 \definecolor{redi}{rgb}{0.8176,0.0078,0.0078}

\begin{document}

\title{ Dynamics of nonlinear excitations of helically confined charges}



\author{A. V.  Zampetaki}
\author{J. Stockhofe}
\affiliation{Zentrum f\"{u}r Optische Quantentechnologien, Universit\"{a}t Hamburg, Luruper Chaussee 149, 22761 Hamburg, Germany}
\author{P. Schmelcher}
\affiliation{Zentrum f\"{u}r Optische Quantentechnologien, Universit\"{a}t Hamburg, Luruper Chaussee 149, 22761 Hamburg, Germany}
\affiliation{The Hamburg Centre for Ultrafast Imaging, Luruper Chaussee 149, 22761 Hamburg, Germany}

\date{\today}

\begin{abstract}
We explore the long-time dynamics of a system of identical charged particles trapped on a closed helix. 
This system has recently been found to exhibit an unconventional deformation of the linear spectrum when tuning the helix radius.
Here we show that the same geometrical parameter can affect significantly also the dynamical behaviour of an initially broad excitation for long times.
In particular, for small values of the radius, the excitation disperses into the whole crystal whereas within a specific narrow regime of larger radii
the excitation self-focuses, assuming finally a  localized form. Beyond this regime, the excitation defocuses and the dispersion gradually increases  again.
We analyze this geometrically controlled nonlinear behaviour using an effective discrete nonlinear Schr\"{o}dinger  model, which allows us among others to
identify a number of breather-like excitations.

\end{abstract}

\maketitle

\begin{center}
 \textbf{I. INTRODUCTION} 
\end{center}

Whereas the harmonic approximation of interactions provides valuable information about the stability and the propagation of small amplitude excitations
in crystals formed by interacting particles, their real-time dynamics as  well as their thermal and transport properties are typically
subject to some degree of nonlinearity \cite{Scott1999}.
Among the most prominent manifestations of such a nonlinearity, are the self-focusing or  self-trapping  \cite{Akhmanov1968, Eilbeck1985, Johansson1995} of initial wave packet excitations
and the existence of non-spreading excitations like breathers and kinks \cite{Sievers1988,Remoissenet1999,Campbell2004}. For discrete systems, 
 a prototype equation incorporating these features is the so-called discrete nonlinear Schr\"{o}dinger (DNLS) equation consisting of a linear (dispersive) coupling  
 and a cubic nonlinear term \cite{Kevrekidis2009}, used to model plenty of systems ranging from coupled optical waveguides \cite{Christodoulides1988,Morandotti1999,Sukhorukov2003}
and Bose-Einstein condensates \cite{Trombettoni1997,Abdullaev2001,Hennig2010} to transport in DNA molecules \cite{Mingaleev1999,Peyrard2004,Koko2012}. The standard spatial arrangement
of sites in most of such one dimensional (1D) studies is that of a straight  equidistant chain in which the coupling (hopping) is restricted to nearest-neighbors (NN).

Non-trivial lattice geometries for 1D discrete nonlinear systems have also been studied and have been found to lead to intriguing new phenomena owing to the interplay
between geometry and nonlinearity. In particular, in curved 1D lattices embedded in a 2D space the bending can act as a trap of excitations 
and induce a symmetry breaking of nontopological solitons \cite{Gaididei2000}. For a 3D space, the helicoidal lattice structure, such as that of DNA molecules,
is found to enhance the existence and stability of discrete breathers \cite{Archilla2001,Sanchez2002}. Furthermore, a curved geometry has been proven  to
 induce nonlinearity in systems where the underlying interactions are harmonic \cite{Kevrekidis2004,Takeno2005}.

In the present work we examine the interplay between nonlinearity and geometry in a system of identical
charged particles, confined on a curved 1D manifold embedded in the 3D space, namely a closed (toroidal) helix. In a previous work \cite{Zampetaki2015}, we have shown that in such a 
system, a tuning of the geometry  controlled by the helix radius, leads to an unconventional deformation of the phononic band structure including a regime of strong degeneracy.
As a consequence, the propagation of small amplitude localized excitations is affected significantly and  a specific geometry exists at which the excitations remain localized
up to long times. A natural question therefore arises, what would be the long time dynamics of a general excitation and whether there is some geometrically controllable degree of
nonlinearity inherent in the system which can alter the propagation characteristics. 

We provide an answer to this question by studying the time propagation of an 
initially broad excitation  on the crystal of charges. We find that for values of the helix radius far from the degeneracy regime, the excitation initially spreads  with 
multiple subsequent revivals due to the closed shape of the  crystal. Within the degeneracy regime, however, the initial excitation  \emph{focuses} in the
course of propagation reaching finally a rather localized state, serving as a hallmark of the existing nonlinearity. 
In order to quantify this nonlinearity we construct an effective DNLS model with  additional nonlocal nonlinear terms \cite{Oster2003,Oster2005}. 
Such a model is found to capture qualitatively well the localization and dispersion features of the original dynamics,
providing a deeper insight into the observed effect. 
Even more, it gives us the opportunity to identify some discrete breather-like excitations at the degenerate geometry, thus adding to the dynamical picture.

The structure of this work is as follows. In Sec. II we describe our system commenting also on its linearized  behaviour. In Sec. III we present 
our results for the  time evolution of an excitation in the crystal for different geometries. In Sec.  IV we construct an effective DNLS model for
our system at the geometries of interest and in Sec. V we use it to identify some breather-like excitations. Finally Sec. VI contains our conclusions.

 \begin{center}
 { \textbf{II. SETUP AND LINEARIZATION}}
\end{center}

\begin{figure}[t!]
\begin{center}
\includegraphics[width=\columnwidth]{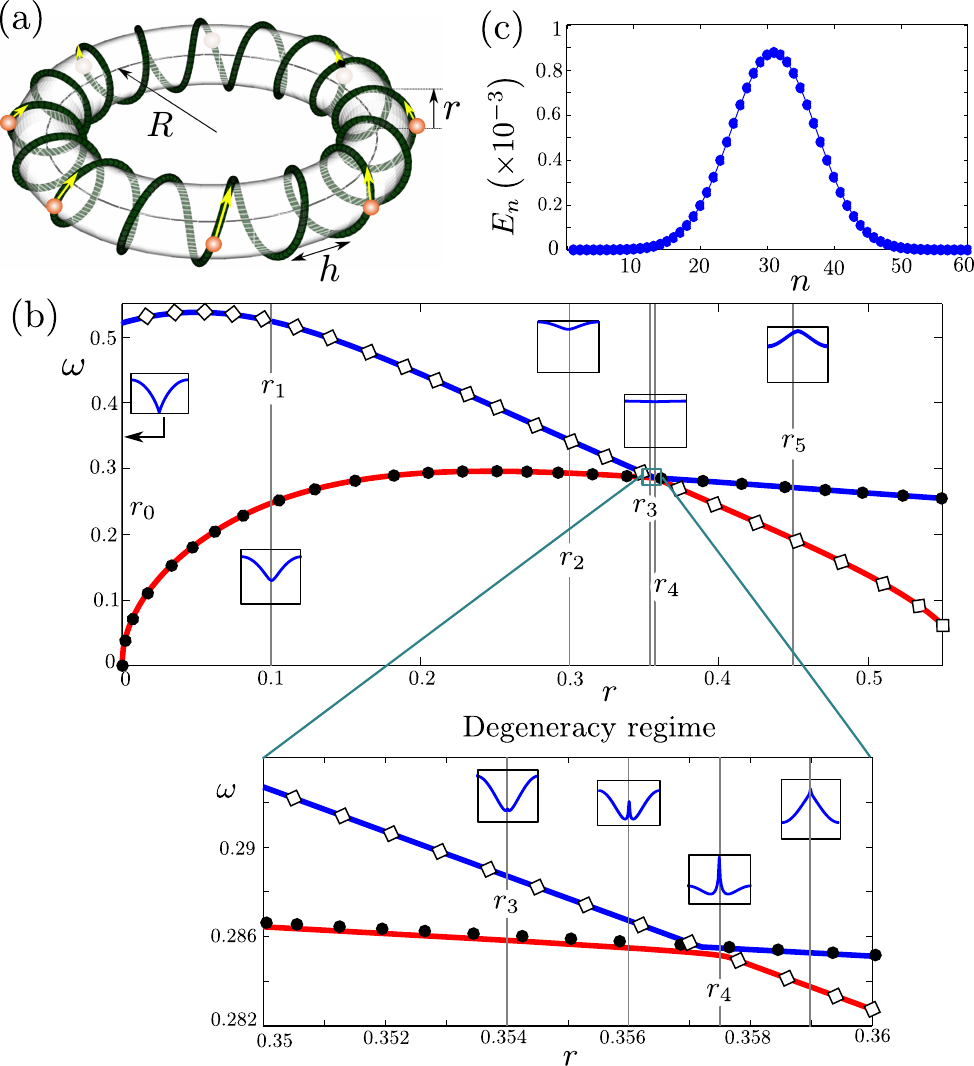}
\end{center}
\caption{\label{tore1} (color online). (a) Equidistant configuration of ions confined on the toroidal helix for $\nu=\frac{1}{2}$ and $N=8$. The yellow arrows indicate the initial velocities
of the particles.
(b) Highest (solid blue line) and lowest (solid red line) frequencies of the linearization spectrum around the 
equilibrium equidistant configuration as a function of $r$ for $N=60$ particles. 
The black dots and the empty diamonds refer to the frequencies corresponding to the center of mass and the out of phase mode respectively. 
The vertical lines mark the radii of the  helix we use in our calculations. 
The small insets depict the form of the respective vibrational band structures $\omega(k)$ at the corresponding values of $r$.
(c) Initial local energy $E_n$ profile as a function of the particle index $n$.}
\end{figure}

 We consider a system of $N$ identical charges of  mass $m_0$ which interact via repulsive Coulomb interactions and are confined 
to move on a 1D toroidal helix, parametrized as
\begin{equation}
 \vec{r} (u)= \begin{pmatrix}
 \left( R+r \cos(u)\right)\cos(au) \\
 \left( R+r \cos(u)\right)\sin(au)\\
 r\sin(u)
\end{pmatrix}, \quad u \in [0, 2 M \pi].
\label{te1}
\end{equation}
In eq. (\ref{te1})  $R$ denotes the major radius of the torus (Fig.\ref{tore1}(a)), $h$ is the helix pitch and $r$ the radius of the helix (minor radius of the torus),
whereas $a=\frac{1}{M}$ stands for the inverse number of windings $M=\frac{2 \pi R}{h}$.
The total effective interaction potential which results from the constrained motion of the charges on the helical manifold reads 
$V(u_1,u_2,\ldots u_N)=\frac{1}{2}\sum_{i,j=1,i \neq j}^{N}\frac{\lambda}{\abs{\vec{r}(u_i)-\vec{r}(u_j)}}$, where $u_j$ denotes the coordinate 
of particle $j$ and $\lambda$ is the coupling constant
characterizing the standard Coulomb interactions. 

 We choose dimensionless units by scaling all our physical quantities (e.g. position $x$, time $t$ and energy $E$)  with $\lambda$, $m_0$ and $2h/\pi$
 as follows 
 
\[\tilde{x}=\frac{x \pi}{2h}, ~\tilde{t}=t\sqrt{\frac{\lambda \pi^3}{8 m_0 h^3}},~ \tilde{E}=\frac{2E h }{\lambda \pi},~\tilde{m}_0=1,~\tilde{\lambda}=1,\]
omitting in the following the tilde for simplicity.
 
  At commensurate fillings, i.e. $M=nN$, $n=1,2,\ldots$ with the  filling factor being $\nu=1/n \leq 1$, it is found that for values of the helix radius $r$
 up to a critical point $r_c$ the ground state configuration of such a system is the equidistant polygonic configuration $u_j^{(0)}=2\left(j-1\right) \pi n$ (Fig. \ref{tore1}(a)). 
 This configuration loses its stability at $r_c$ undergoing a zig-zag bifurcation \cite{Zampetaki2015}.
 
 We focus in this work on the dynamical behaviour of charged particles confined on the toroidal helix in the region $r<r_c$, where the ground state 
is still the polygonic one. We have shown in \cite{Zampetaki2015} that in such a region the linear spectrum of the system changes dramatically with tuning
the radius of the helix $r$, a fact that crucially affects the
 propagation of small amplitude localized excitations. Specifically, it was found that the width of the linear spectrum decreases as one approaches a point $r_d$ 
 of strong degeneracy from below and increases again beyond that point, while interchanging the character between the eigenmodes corresponding  to the highest and the lowest
 frequencies (Fig. \ref{tore1}(b)). In fact, since the degeneracy is not complete, it is better to refer to  a  degeneracy regime
 within which  the inversion of the spectrum is gradually achieved while its width remains small (Fig. \ref{tore1}(b) (inset)).
 We consider in this work six different geometries, each corresponding to 
 a different value of $r$, covering all the regions with a qualitatively different linear spectrum (Fig. \ref{tore1}(b)) from the ring limit ($r_0=0$) to the degeneracy ($r_3,r_4$)
 and the inversion ($r_5$) regime. We focus on the case of half-filling  $\nu=\frac{1}{2}$ for $N=60$  particles.
 
 \begin{center}
 \textbf{III. TIME PROPAGATION OF A GAUSSIAN EXCITATION} 
\end{center}
 
 In this section we present and discuss the dynamical response of our system to an initial excitation. Although the physical results are
 in principle independent of the exact character of this excitation and the means used for its quantification, the determination of both is essential 
 for the illustration and the theoretical description of our findings.

 Dealing with classical systems and seeking an excitation measure whose
 total amount is conserved in time, the natural choice  is  a, to be defined, energy distribution associated with each  particle, referred to hereafter as
 local energy $E_n$. Whereas the kinetic energy $K$ consists of parts allocated to each individual particle,  the potential energy
 cannot be uniquely partitioned, yielding different definitions of local energy \cite{Allen1998,Sarmiento1999,Hennig2007}. Aiming for them to be 
 strictly positive for
all possible excitations   (a considerably non-trivial requirement for systems with Coulomb interactions),
we define our local energies $E_n$ in a rather unconventional way, focusing on a positive decomposition of the harmonic interaction term \cite{Allen1998}.
 Our complete definition and a more detailed discussion of local energies are provided in the Appendix.
 
We start with an initially broad excitation of a Gaussian profile in terms of local energies (Fig. \ref{tore1}(c)). Since the local energies $E_n$ 
depend trivially on the particles' velocities (contrary to what is the case for the particles' positions),
the most straightforward way to obtain such a Gaussian local energy profile is by exciting the particles with a suitable velocity distribution.
Of course, for a given local energy profile the magnitude of such a velocity distribution  can be uniquely determined, but there is 
a freedom in the direction of the velocity for each particle. We choose here  all the velocities to point in the positive direction (Fig. \ref{tore1}(a)).
%

\begin{figure*}[htbp]
\centering
\includegraphics[width=1.8\columnwidth]{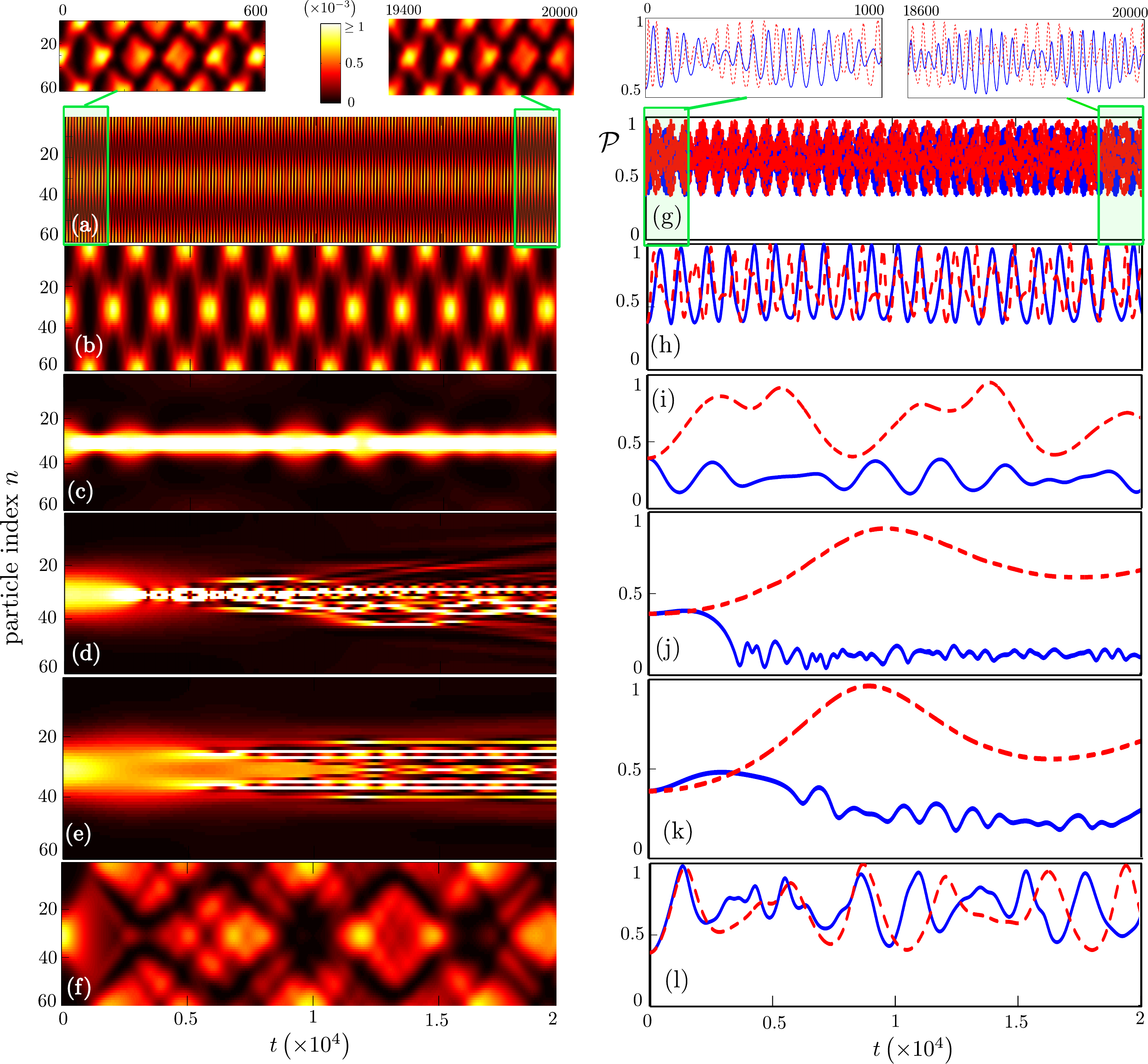}
\caption{\label{prop1} (color online) (a)-(f) Time evolution of the initial Gaussian excitation presented in Fig. \ref{tore1}(c) for $N=60$, $\nu=1/2$ for increasing $r$
corresponding to the points (a) $r_0$, (b) $r_1$, (c) $r_2$, (d) $r_3$, (e) $r_4$ and  (f) $r_5$ marked in Fig.\ref{tore1}(b). Colors encode the values of local energy $E_n$ for each particle $n$ and
time $t$. For the same values of $r$ panels (g)-(l) depict the time evolution of the normalized participation ratio  $\mathcal{P}$ of the excitation.
The solid blue lines are the results from our numerical simulations corresponding to (a)-(f), whereas the dashed red lines correspond to the results for
a harmonic approximation of the potential.
}
\end{figure*}

The ensuing dynamics is shown in (Fig. \ref{prop1}(a)-(f)). Obviously, the time evolution of the Gaussian excitation possesses a drastic dependence  
on the geometry, controlled by the helix radius $r$.
In particular, for $r<r_d$ (Fig. \ref{prop1}(a),(b)) the excitation spreads into the whole crystal and refocuses almost periodically at the time instants when the left and the right propagating
parts of the excitation meet and superimpose at the diametrically opposite point of the  closed helix. As discussed in \cite{Zampetaki2015} the spreading
velocity decreases as $r$ is increased,
following the width of the linear spectrum  (Fig. \ref{tore1}(b)).
As the width becomes smaller the features of the time evolution alter significantly. 
 Already at the point $r_2$ the excitation does not spread any more into the crystal, but it alternately focuses and again defocuses to its initial shape (Fig. \ref{prop1}(c)). Even more surprisingly,
within the degeneracy region (Fig. \ref{prop1}(d),(e)) the initial excitation undergoes a focusing after some time scale $t_F$ ($t_F\approx 4000$ for (d), $t_F\approx 6000$
for (e)). Subsequently, the wave packet loses its smooth envelope and fragments into a number of highly localized excitations.
 Depending on $r$, the routes towards such a localized state can  be different,
with the wave packet evolving initially one central peak (Fig. \ref{prop1}(d)) or two side peaks (Fig. \ref{prop1}(e)).
Another interesting feature  within the degeneracy regime is that the reflection symmetry of the initial excitation profile can break in the course of propagation, attaining after
some time a significantly asymmetric form (Fig. \ref{prop1}(d)). The direction of the asymmetry depends on the direction of the 
initial velocities of the particles.
Beyond the degeneracy (Fig. \ref{prop1}(f)) the spreading of the excitation  into the crystal reappears  with a periodic refocusing but the propagation pattern is much 
different owing to the inverted form of the vibrational band structure.

In order to quantify the degree of focusing or localization of the excitation,
we examine the time dependence of the  normalized participation ratio 
\begin{equation}
\mathcal{P}= \frac{1}{N}\frac{\left(\sum_{n=1}^{N} E_n\right)^2}{\sum_{n=1}^{N} E_n^2}. \label{prat}
\end{equation}

Evidently, this quantity can take values between $1/N$ and $1$, with $\mathcal{P}=1$ signifying the case of a completely extended excitation
where the energy is equipartitioned between all the particles and $\mathcal{P}=1/N$ marking the opposite case of a fully localized excitation in a 
single particle. Note that the local energies $E_n$ in (\ref{prat}) should be non-negative for the definition to make sense.

Our results, presented in Figs. \ref{prop1}(g)-(l), support our discussion above. Especially the focusing of the excitation after $t_F$ is
evident in Figs. \ref{prop1}(j)-(k).  However, the subsequent drop in $\mathcal{P}$ is much stronger
in Fig. \ref{prop1}(j) than in Fig. \ref{prop1}(k), in line with the observation that at  $r=r_3$ the final localized 
state consists of less excited particles (Fig. \ref{prop1}(d)) than at $r=r_4$ (Fig. \ref{prop1}(e)).

In \cite{Zampetaki2015} it was demonstrated that a  
small localized initial excitation does not spread significantly for short times at the degeneracy point, in contrast to
the behaviour for other geometries.
This fact was understood solely by an inspection of the linearization spectrum. 
Here, however, the situation is  different. Not only the complete absence of spreading, but especially the  
existence of self-focusing calls for an account of the underlying nonlinearity. 
This is further emphasized and supported by  Figs. \ref{prop1}(g)-(l) where the results of
the propagation within the harmonic approximation of the potential are also displayed. As long as the total amplitude of the  initial excitation 
is small enough, the harmonic approximation  works well.
As the amplitude is increased this approximation will start to fail, and nonlinear effects are expected to show up.
The results of the present work suggest that except for the amplitude, also the geometry allows to control the importance of 
the nonlinearity. 
Specifically, for the given amplitude and for geometries far from the degeneracy regime (Figs. \ref{prop1}(g),(h),(l))
the harmonic approximation  qualitatively reproduces the  exact  time evolution of the participation ratio,
although, as should be expected, there are quantitative deviations.
In contrast, close to and within the degeneracy regime
the harmonic approximation fails completely, predicting a spreading and an extended form of the excitation, instead of localization. This makes it clear 
that regarding the focusing, we  indeed encounter a nonlinear phenomenon.

\begin{figure}[htbp]
\begin{center}
\includegraphics[width=6cm]{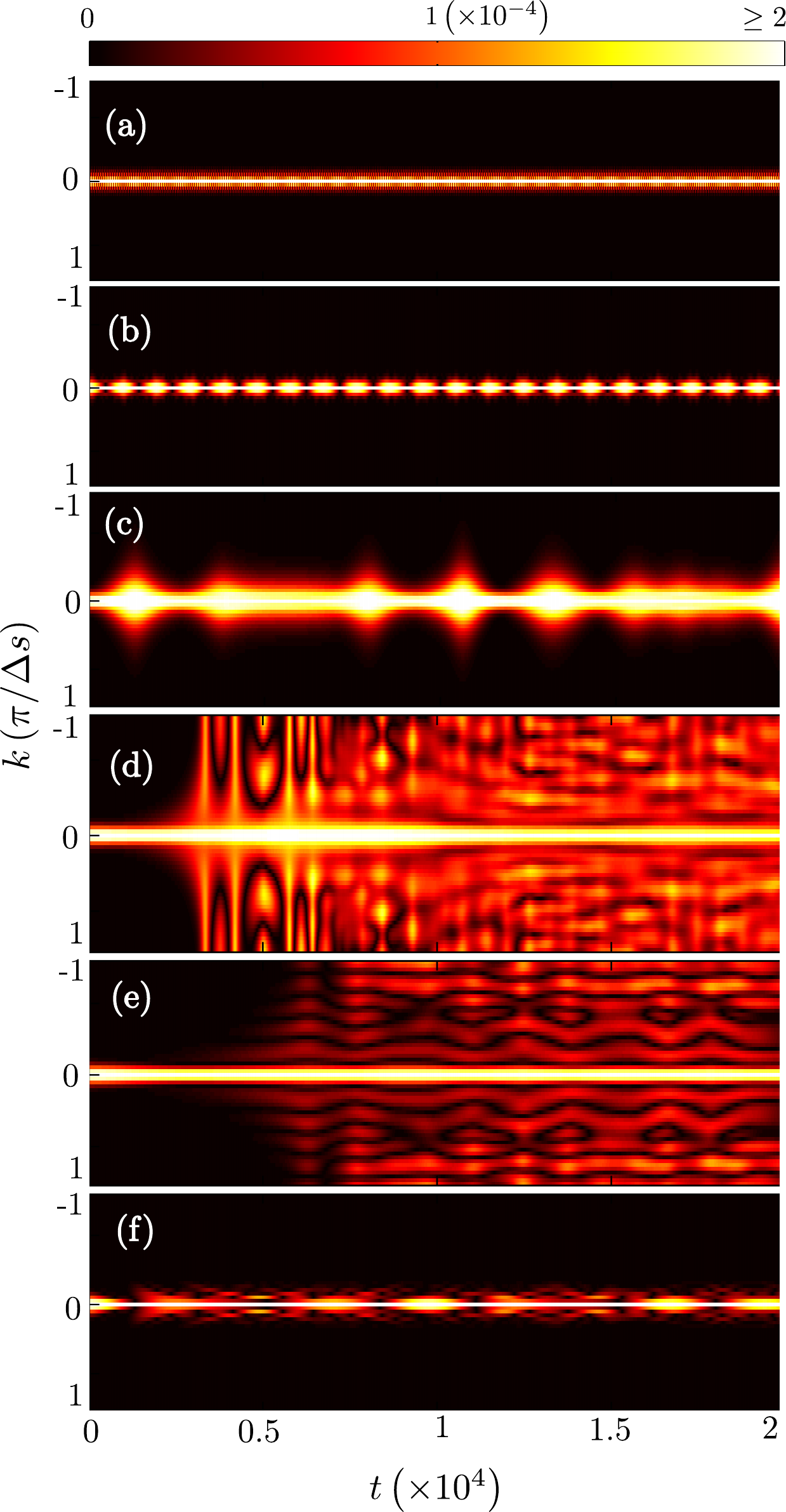}
\end{center}
\caption{\label{fou1} (color online) Absolute value of the discrete Fourier transform 
of the local energy excitation profile as a function of time and wave number $k$. Subfigures (a) to (f) correspond to increasing $r$, ranging from $r_0$ to $r_5$
as indicated in Fig. \ref{tore1} (b).}
\end{figure}

 Complementary information about the spreading or  localization of the Gaussian excitation can be obtained from its time evolution in the 
 reciprocal space, i.e. in terms of the wave numbers $k\in\left[-\frac{\pi}{\Delta s},\frac{\pi}{\Delta s}\right]$, where $\Delta s$ denotes
 the arc length interparticle distance of the equidistant ground state configuration. 
 To this extent, we examine how the discrete Fourier transform of the local energy  profile evolves in the reciprocal space
 (Figs. \ref{fou1}(a)-(f)).
 The initial excitation, being overall extended in the coordinate space (Fig. \ref{tore1}(c)), appears to be
 rather localized around the wave vector $k=0$ in the $k$-space. It remains localized as expected
for $r<r_d$ (Figs. \ref{fou1}(a),(b)), but when approaching $r_d$ more wave vectors in the vicinity of $k=0$ become excited (Fig. \ref{fou1}(c)). At the degenerate geometries
(Figs. \ref{fou1}(d),(e)) the excitation expands rapidly in the reciprocal space,
populating after the characteristic time  $t_F$ almost all the wave numbers $k$. The space localized solution
consists therefore of most of the $k$ modes, with the population of the initial $k=0$  mode being dominant.
At $r>r_d$ (Figs. \ref{fou1}(f)) the excitation remains, as for small $r$,  in the narrow vicinity of the $k=0$ mode.

Before proceeding with our study of the nonlinear behaviour, let us note that 
for the results presented here we have used a rather small initial excitation with a total energy  of
the order of $1\%$ of the ground state energy per particle $E_{GS}/N$. For larger amplitudes the self-focusing 
can occur also for smaller radii, i.e.  at a greater distance from the degeneracy region.
 \begin{center}
 \textbf{IV. EFFECTIVE NONLINEAR MODEL} 
\end{center}

The dynamics analyzed in the previous section is characterized by a self-focusing process of excitations for the case of degenerate geometries
and therefore suggests  
a prominent role of the nonlinearity. We aim in this section to identify and quantify the leading nonlinear terms as well as to derive a  DNLS effective model 
in the region of degeneracy.
As a result we will, among others, gain insight into the excitation amplitudes and the time scale $t_F$ for localization.

 \begin{center}
 \textbf{A. The dominant nonlinear terms} 
\end{center}

Since the initial excitation is small enough we attempt to
identify the dominant anharmonic terms  by expanding the potential around the equilibrium configuration $\{u^{(0)}\}$ up to fourth order.
It is advantageous to  do so in the arc length parametrization $s$ so that the final Euler-Lagrange equations of motion and particularly the kinetic terms would
assume the standard form \cite{Zampetaki2013}. 
To this purpose we calculate the matrices
\begin{equation}
 H_{ij}=\left.\frac{\partial^2 V}{\partial s_i \partial s_j}\right|_{\{s^{(0)}\}},~G_{ij}=\left.\frac{\partial^3 V}{\partial s_i^2 \partial s_j}\right|_{\{s^{(0)}\}}
\end{equation}
corresponding to the Hessian and the matrix of the third derivatives respectively. Terms involving more than two different positions are 
equal to zero, since the total potential $V$ is a sum of exclusively two-body potential terms.

For the fourth derivative terms we define two further matrices as
\begin{equation}
 M_{ij}=\left.\frac{\partial^4 V}{\partial s_i^2 \partial s_j^2}\right|_{\{s^{(0)}\}},~Q_{ij}=\left.\frac{\partial^4 V}{\partial s_i^3 \partial s_j}\right|_{\{s^{(0)}\}}
\end{equation}
for $i\neq j$ and \[M_{ii}=Q_{ii}=\frac{1}{2}\left.\frac{\partial^4 V}{\partial s_i^4}\right|_{\{s^{(0)}\}},\]
 separating the derivatives of the same order in $s_i$, $s_j$ from those with a different order   
and splitting the diagonal derivative terms (involving differentiation with respect to a single position) in two.
The calculation of the above derivatives in the arc length parametrization can be carried out by using the respective derivatives in the $u$ coordinate space
as well as the known relations for derivatives of  inverse functions.
Under these considerations and denoting $s_j-s_j^{(0)}=x_j$ the potential reads
\begin{align}
V& \approx E_{GS}+\frac{1}{2}\sum_{i,j=1}^N x_i H_{ij} x_j+\frac{1}{6}\sum_{i,j=1}^N x_i^2 G_{ij} x_j\nonumber \\ 
&+\frac{1}{24}\sum_{i,j=1}^N x_i^2 M_{ij} x_j^2
+\frac{1}{24}\sum_{i,j=1}^N x_i^3 Q_{ij} x_j
\end{align}
leading to the equations of motion
\begin{align}
\ddot{x}_n&=-\sum_{ j\neq n} H_{nj} x_j-H_{nn} x_n-\frac{1}{3}\sum_{ j \neq n} x_n G_{nj} x_j\nonumber \\ 
&-\frac{1}{6}\sum_{ j \neq n}  G_{jn} x_j^2-\frac{1}{6}\sum_{ j \neq n} x_n M_{nj} x_j^2-\frac{1}{3} M_{nn}x_n^3\nonumber \\ 
&-\frac{1}{8}\sum_{j\neq n} x_n^2 Q_{nj} x_j-\frac{1}{24}\sum_{ j\neq n}  Q_{jn} x_j^3,\label{eom}
\end{align}
which consist of harmonic, quadratic and third order nonlinear terms. All the matrices appearing in these equations are symmetric
except for the matrix $G_{ij}$, relating to the quadratic force terms, which is fully antisymmetric.
This can affect significantly the symmetry of the expected solutions. If the quadratic nonlinear force terms are ignored then 
the equations of motion (\ref{eom}) possess the symmetry $x_n \rightarrow -x_n$ which also results  in symmetric excitations keeping their 
symmetry in the course of propagation. The quadratic force terms, however, break the reflection symmetry and allow for an 
asymmetric evolution of initially symmetric excitations as
the one observed in Fig. \ref{prop1}(d).

Let us mention at this point that in the ring limit $r=r_0=0$, where a separation of the center of mass holds, the matrices involved
in eq. (\ref{eom}) are not independent but obey the relations
\[H_{nn}=-\sum_{\substack{ j\neq n}} H_{nj},~M_{nn}=-\frac{1}{8}\sum_{\substack{ j\neq n}} Q_{nj},~M_{nj}=-\frac{3}{4}Q_{nj}\]
yielding
\begin{align}
 \ddot{x}_n&=-\sum_{ j\neq n} H_{nj} (x_j-x_n)-\frac{1}{6}\sum_{ j\neq n} G_{jn} (x_j-x_n)^2 \nonumber \\
 &-\frac{1}{24}\sum_{ j\neq n} Q_{nj} (x_j-x_n)^3 
\end{align}
which for only nearest neighbor (NN) couplings  leads to a Fermi-Pasta-Ulam  kind of equations of motion \cite{Fermi1995},
with both quadratic and cubic nonlinear interactions. 

 Before proceeding, we note that the matrix elements  $H_{ij},G_{ij},M_{ij},Q_{ij}$  depend,
due to the symmetry of the ground state configuration, only on the index difference  $m=i-j$. Therefore, when referring to these elements in the following  we will 
use  the notation $H_{m},G_{m},M_{m},Q_{m}$.

 \begin{center}
 \textbf{B. The DNLS model} 
\end{center}

\begin{figure}[t!]
\begin{center}
\def\svgwidth{\columnwidth}
\begingroup%
  \makeatletter%
  \providecommand\color[2][]{%
    \errmessage{(Inkscape) Color is used for the text in Inkscape, but the package 'color.sty' is not loaded}%
    \renewcommand\color[2][]{}%
  }%
  \providecommand\transparent[1]{%
    \errmessage{(Inkscape) Transparency is used (non-zero) for the text in Inkscape, but the package 'transparent.sty' is not loaded}%
    \renewcommand\transparent[1]{}%
  }%
  \providecommand\rotatebox[2]{#2}%
  \ifx\svgwidth\undefined%
    \setlength{\unitlength}{781.7830113bp}%
    \ifx\svgscale\undefined%
      \relax%
    \else%
      \setlength{\unitlength}{\unitlength * \real{\svgscale}}%
    \fi%
  \else%
    \setlength{\unitlength}{\svgwidth}%
  \fi%
  \global\let\svgwidth\undefined%
  \global\let\svgscale\undefined%
  \makeatother%
  \begin{picture}(1,0.97716942)%
  
    \put(0,0){\includegraphics[width=\unitlength]{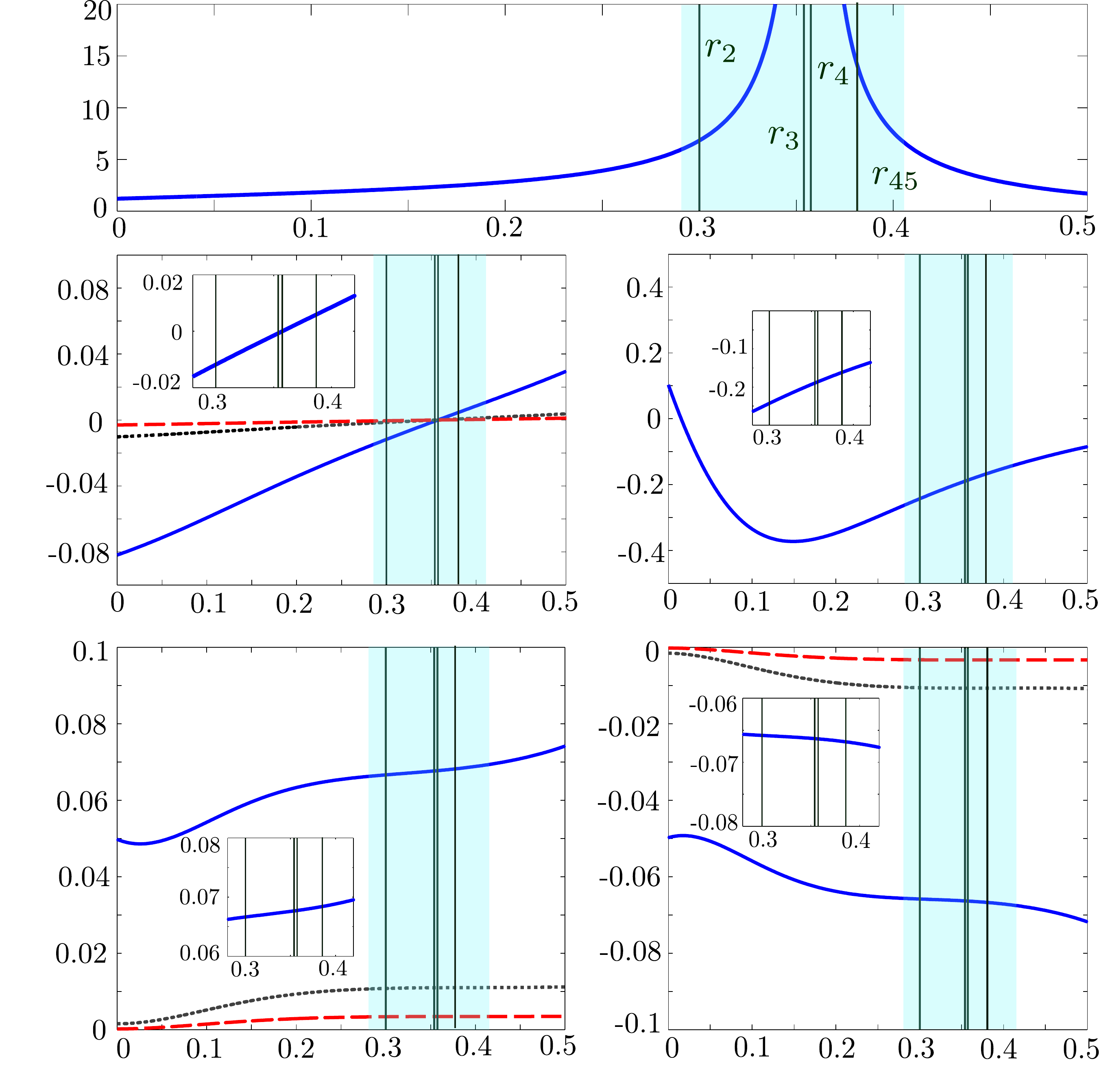}}%
    \put(-0.0142444,0.87419252){\color[rgb]{0,0,0}\makebox(0,0)[lb]{\smash{$\frac{\abs{H_{0}}}{2\abs{H_{1}}}$}}}%
    \put(0.00688392,0.94930875){\color[rgb]{0,0,0}\makebox(0,0)[lb]{\smash{\scalebox{0.99}{(a)}}}}%
    \put(0.00078392,0.69275964){\color[rgb]{0,0,0}\makebox(0,0)[lb]{\smash{\scalebox{0.99}{(b)}}}}%
    \put(0.52198903,0.69275964){\color[rgb]{0,0,0}\makebox(0,0)[lb]{\smash{\scalebox{0.99}{(c)}}}}%
    \put(0.00688392,0.367869){\color[rgb]{0,0,0}\makebox(0,0)[lb]{\smash{\scalebox{0.99}{(d)}}}}%
    \put(0.53198903,0.367869){\color[rgb]{0,0,0}\makebox(0,0)[lb]{\smash{\scalebox{0.99}{(e)}}}}%
    \put(0.0222631,0.60036458){\color[rgb]{0,0,0}\makebox(0,0)[lb]{\smash{\scalebox{0.99}{$A_l$}}}}%
    \put(0.5427203,0.60012191){\color[rgb]{0,0,0}\makebox(0,0)[lb]{\smash{\scalebox{0.99}{$B$}}}}%
    \put(0.0222631,0.28802125){\color[rgb]{0,0,0}\makebox(0,0)[lb]{\smash{\scalebox{0.99}{$C_l$}}}}%
    \put(0.5327203,0.2868113){\color[rgb]{0,0,0}\makebox(0,0)[lb]{\smash{\scalebox{0.99}{$D_l$}}}}%
    \put(0.30266928,0.00135076){\color[rgb]{0,0,0}\makebox(0,0)[lb]{\smash{\scalebox{0.99}{$r$}}}}%
    \put(0.7857753,0.0044816){\color[rgb]{0,0,0}\makebox(0,0)[lb]{\smash{\scalebox{0.99}{$r$}}}}%
    \put(0.78672286,0.41745206){\color[rgb]{0,0,0}\makebox(0,0)[lb]{\smash{\scalebox{0.99}{$r$}}}}%
    \put(0.30188423,0.41583845){\color[rgb]{0,0,0}\makebox(0,0)[lb]{\smash{\scalebox{0.99}{$r$}}}}%
    \put(0.54525104,0.77058797){\color[rgb]{0,0,0}\makebox(0,0)[lb]{\smash{\scalebox{0.99}{$r$}}}}%
  \end{picture}%
\endgroup%
\end{center}
\caption{\label{coef1} (color online). (a) The ratio between the diagonal and off diagonal elements of the Hessian $\frac{\abs{H_{0}}}{2\abs{H_{1}}}$ as a function of $r$.
(b)-(e) Coefficients $A_l, B, C_l, D_l$ defined in eq. (\ref{coe2}) as a function of $r$. The solid blue lines depict the coefficient values for $l=1$, whereas the 
black doted lines for $l=2$ and the red dashed ones for $l=3$. The cyan (shaded) area marks the region of applicability of the RWA as derived from the panel (a).
The vertical  lines correspond to the values of $r$ we study with the DNLS model. Note that  three of them namely $r_2$,$r_3$ and $r_4$ are the same as those 
studied in the previous section (Fig. \ref{tore1}(b)) whereas $r_{45}$ lies between $r_4$ and $r_5$.
The insets provide a zoom into the region of interest.}
\end{figure}
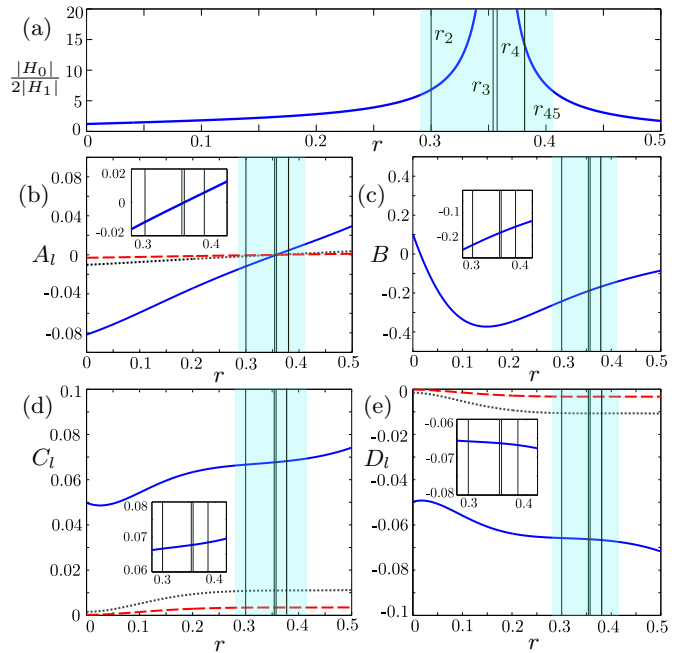

The  effect of  localization of the initial wave packet occurs, as we have observed in Sec. III, in the regime 
close to degeneracy. There, the  diagonal terms of the  Hessian provide the dominant contribution to the linear spectrum 
as the off-diagonal ones are very  small.
Since $\abs{H_{0}} \gg 2\abs{H_{1}}$ (Fig. \ref{coef1}(a)),  one can use in that regime the so-called rotating wave approximation (RWA)
\cite{Kivshar1992},  assuming that the position coordinate
can be described as 
\begin{equation}
 x_n(t)=\Psi_n(t) e^{-i\omega_0 t}+\Psi_n^*(t) e^{i\omega_0 t},\label{ans1}
\end{equation}
with $\Psi_n(t)$ a slowly varying amplitude and $\omega_0^2=H_{0}$ yielding a fast oscillating phase $e^{\pm i \omega_0 t}$. Apart 
from the requirement of a weak dispersion,  a condition for a sufficiently weak nonlinearity has also to be satisfied \cite{Kivshar1992},
namely $\abs{H_{0}} \gg \frac{1}{3}\abs{M_{0}}\max (x_i(0))^2$ with $\max (x_i(0))$ the maximum initial displacement (or respectively momentum)
of a single particle. This  criterion is satisfied as well in our case, since the initial conditions we have used lead to  $\abs{H_{0}}$
of about $100$ times larger than $\frac{1}{3}\abs{M_{0}}\max (x_i(0))^2$.

Using the ansatz (\ref{ans1}) and making the assumption $\abs{\frac{d\Psi_n}{dt}} \ll \omega_0 \abs{\Psi_n}$,
as well as neglecting the rapidly oscillating terms with frequency $3 \omega_0$,
 the equations of motion (\ref{eom}) acquire the form
\begin{align}
&i\dot{\Psi}_n=\sum_{l=1}^{N/2} A_l (\Psi_{n-l}+\Psi_{n+l})+B\abs{\Psi_n}^2\Psi_n \nonumber \\
&+\sum_{l=1}^{N/2} C_l \left[2\Psi_n\left(\abs{\Psi_{n-l}}^2+\abs{\Psi_{n+l}}^2\right)+\Psi_n^*(\Psi_{n-l}^2+\Psi_{n+l}^2)\right]\nonumber \\ 
&+\sum_{l=1}^{N/2} D_l \left[2\abs{\Psi_n}^2\left(\Psi_{n-l}+\Psi_{n+l}\right)+\Psi_n^2\left(\Psi_{n-l}^*+\Psi_{n+l}^*\right)\right.\nonumber \\
&\left.+\abs{\Psi_{n-l}}^2\Psi_{n-l}+\abs{\Psi_{n+l}}^2\Psi_{n+l}\right] \nonumber \\
&+\sum_{l=1}^{N/2} F_l \left[\left(2\Psi_n\left(\Psi_{n+l}-\Psi_{n-l}\right)-\Psi_{n+l}^2+\Psi_{n-l}^2\right)  e^{-i\omega_0 t}\right.\nonumber \\
&\left.+\left(2\Psi_n\left(\Psi_{n+l}^*-\Psi_{n-l}^*\right)-\abs{\Psi_{n+l}}^2+\abs{\Psi_{n-l}}^2\right) e^{i\omega_0 t}\right] \label{eqf1}
\end{align}
with
\begin{align}
A_l&=\frac{H_{l}}{2 \omega_0},~B=\frac{M_{0}}{2 \omega_0},~ C_l=\frac{M_{l}}{12 \omega_0}\nonumber \\
D_l&=\frac{Q_{l}}{16 \omega_0},~ F_l=\frac{G_{l}}{12 \omega_0}.\label{coe2}
\end{align}
The last term in eq. (\ref{eqf1}) comes from the quadratic nonlinear force term which is responsible for generating second harmonics and breaking the 
reflection symmetry of the equations. Within the RWA this term is also neglected due to its fast oscillation in time in comparison to the other 
slowly varying terms.
Of course such an approximation limits our consideration to the general focusing/defocusing behaviour of the wave packet excitation,
not accounting  for secondary propagation features such as the exact shape of the excitation pulse during the time evolution.

The actual numerical values of the remaining coefficients $A_l,B, C_l,D_l$ for our system are shown
in Figs. \ref{coef1}(b)-(e) as a function of $r$. All the coefficients in the region of validity of our approximation have a definite sign 
except for $A_l$ which changes sign within the degeneracy regime.  It is also obvious
that  an increasing index $l$ leads to coefficients closer and closer to zero, with even the coefficients of $l=2$ being already very small.
Since the major contribution  stems  from $l=1$, we can proceed performing a NN approximation which yields
\begin{align}\label{dnls1}
&i\dot{\Psi}_n= A (\Psi_{n-1}+\Psi_{n+1})+B\abs{\Psi_n}^2\Psi_n \nonumber \\
& +C \left[2\Psi_n\left(\abs{\Psi_{n-1}}^2+\abs{\Psi_{n+1}}^2\right)+\Psi_n^*(\Psi_{n-1}^2+\Psi_{n+1}^2)\right]\nonumber \\ 
&+ D \left[2\abs{\Psi_n}^2\left(\Psi_{n-1}+\Psi_{n+1}\right)+\Psi_n^2\left(\Psi_{n-1}^*+\Psi_{n+1}^*\right)\right.\nonumber \\
&\left.+\abs{\Psi_{n-1}}^2\Psi_{n-1}+\abs{\Psi_{n+1}}^2\Psi_{n+1}\right] 
\end{align}
where we have omitted the index $1$ from the coefficients for clarity. Equation (\ref{dnls1}) can be recognized as a DNLS equation 
with  additional nonlinear couplings studied in the literature \cite{Oster2003,Oster2005} and will be the point of reference in the following discussion.

\begin{figure}[htbp]
\begin{center}
\includegraphics[width=8cm]{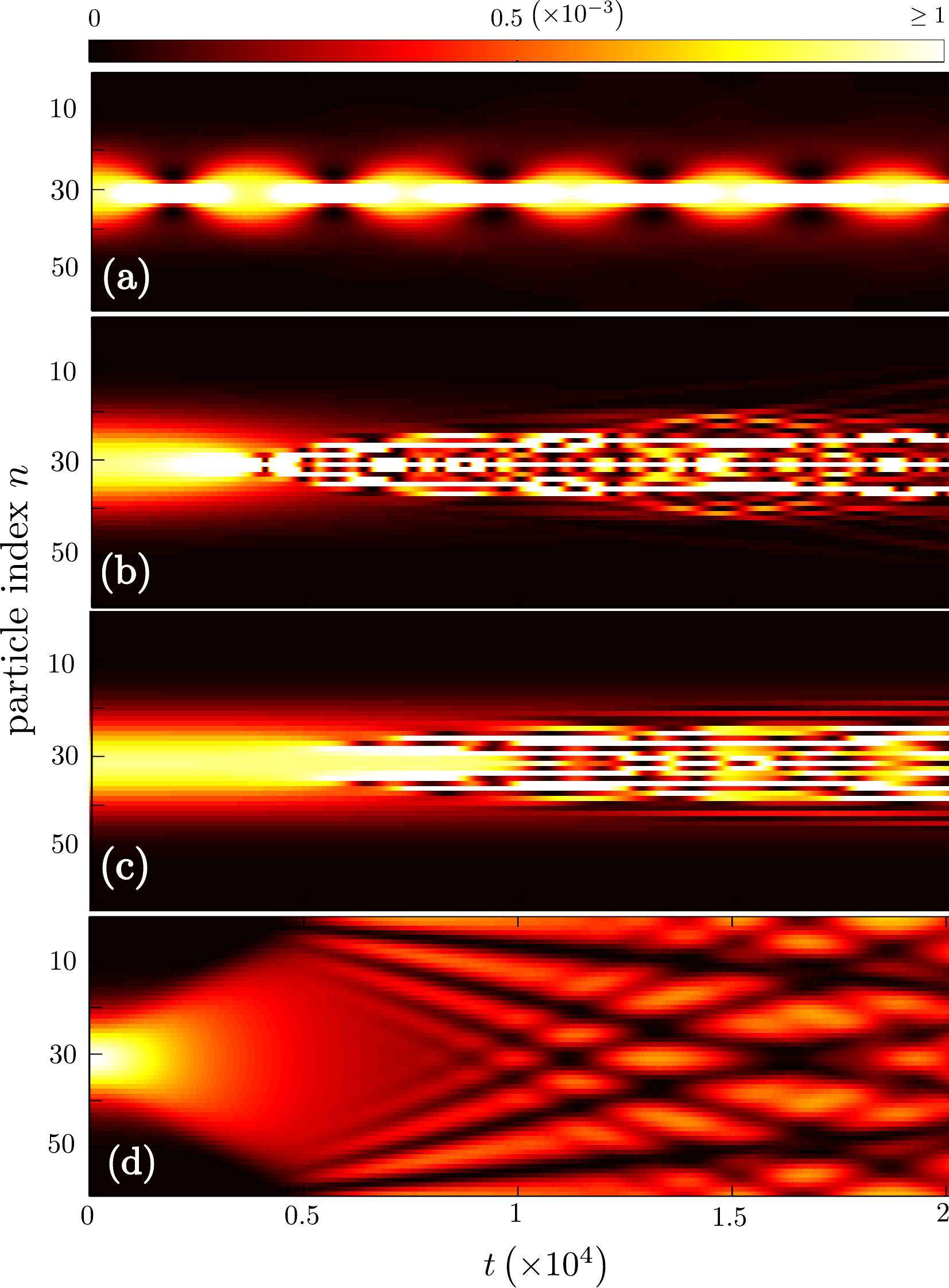}
\end{center}
\caption{\label{fnsch1} (color online). (a)-(d) Time propagation of the initial Gaussian excitation of Fig. \ref{tore1}(c) under the DNLS equation (\ref{dnls1})  for  increasing $r$
corresponding to the points (a) $r_2$, (b) $r_3$, (c) $r_4$ and (d) $r_{45}$ marked in Fig. \ref{coef1}(a).
Colors encode the values of local energy $E_n=2 \omega_0^2 \abs{\Psi_n}^2$ for each particle $n$ and
time $t$.}
\end{figure}

In order to check the validity of our approximation we propagate the  Gaussian excitation illustrated in Fig. \ref{tore1}(c) according to  eq. (\ref{dnls1}).
 Note that the quantity $\Psi$ is by definition complex, with its real part  at $t=0$ relating to the displacement $\textrm{Re}(\Psi_n(0))=\frac{x_n(0)}{2}$
(eq. (\ref{ans1})) 
and its imaginary part to the momentum $\textrm{Im}(\Psi_n(0))=\frac{p_n(0)}{2 \omega_0}$  (eq. (\ref{ans1}) within the RWA),
both in the arc length parametrization.  Furthermore, the local energy 
$E_n=\frac{1}{2}(\omega_0^2 x_n^2+p_n^2)$ can be expressed as  $E_n=2 \omega_0^2 \abs{\Psi_n}^2$, the conservation of the total energy being thus linked directly with
the conservation of $\sum_n \abs{\Psi_n}^2$. A comparison between Figs. \ref{fnsch1}(a)-(c) and Figs. \ref{prop1}(c)-(e) makes it clear that the NN 
DNLS model captures qualitatively very well the features of the exact excitation propagation, exhibiting focusing and defocusing at $r_2$ (Fig.  \ref{fnsch1}(a))
and localization after a certain time for $r$ within the degeneracy region (Figs.  \ref{fnsch1}(b),(c)).
Beyond the point $r_4$, but still very close to the degeneracy 
(Fig. \ref{coef1}(a)), the excitation, although keeping its shape for  a long time, eventually disperses into the whole crystal (Fig.  \ref{fnsch1}(d)).
Note that the profile of the excitation remains reflection symmetric (with respect to the central particle) in the  course 
of propagation for all geometries (Figs.  \ref{fnsch1}(a)-(d)), in contrast to what is observed in (Fig.  \ref{prop1}(d)), which we attribute primarily to the neglected quadratic
nonlinear term.

Most of these results  can already be  inferred  from an inspection of eq. (\ref{dnls1}) and the parameter $A(r)$ as shown in Fig.  \ref{coef1}(b).
When increasing the radius, $A$ changes its sign from negative (same sign as $B$) to positive.
 Correspondingly, with increasing $r$ the effective nonlinearity changes from attractive (hopping $A$ and nonlinearity $B$ of the same sign), leading to self-focusing dynamics (Figs. \ref{fnsch1}(a)-(c)),
to repulsive (hopping $A$ and nonlinearity $B$ of opposite sign), leading to defocusing (Fig. \ref{fnsch1}(d)).
 
 The change of the magnitude of the parameter $A$  affects also the degree and the time-scale of the wave-packet localization.
 A small hopping term is expected to slow down the dynamical behaviour of the system since it accounts for  reduced mobility. A decrease in its magnitude
 leads also to an enhancement of the general nonlinear behaviour since it shifts the weight of the dynamics to the nonlinear terms. 
 On a more formal level, these statements can be justified within a multiple scale analysis \cite{Bender1978}.
 In fact, one can arrive at the RWA by assuming that all the terms of eq. (\ref{eom}) are much smaller than the $H_{nn} x_n$ term. Introducing the small parameter $\epsilon= \abs{A}$,
 due to the weak coupling and weak nonlinearity conditions  a slow time scale can be defined $\tilde{t}=\epsilon t=\abs{A} t$, which is characteristic 
 for the time evolution of $\Psi_n$. A further rescaling of $\Psi_n$ with $\sqrt{\abs{\frac{B}{A}}}$ leads to the standard form of our DNLS \cite{Oster2003,Oster2005}
 with $\tilde{A}=\rm{sgn}(A)$, $\tilde{B}=\rm{sgn}\left(\frac{B}{A}\right)$,
 $\tilde{C}=\frac{C}{B}$, $\tilde{D}=\frac{D}{B}$ and $\tilde{\Psi}_n=\sqrt{\abs{\frac{B}{A}}} \Psi_n$ in which the strength of  the 
 nonlinearity is  primarily controlled by the amplitude of the initial excitation $\tilde{\Psi}_n(0)$. 
 In the degeneracy regime, due to the factor  $\sqrt{\abs{\frac{B}{A}}}$, this is about one order of magnitude larger than its value for other parameter values of $r$
resulting in the observed strong localization of the excitation. In the same regime the time scale of the time evolution becomes very large, scaling as $\frac{1}{\abs{A}}$
and yielding in our case a characteristic time $t_F$ of the order of $10^3$. 

Before closing this section, let us note that in the limit  $\abs{A} \rightarrow 0$ the NN neighbor 
approximation gradually fails (Fig. \ref{coef1}(b)) and the dispersion coefficients $A_l$ from all the neighbors have to be taken into account. 
Furthermore, at long times, after the localization the weak nonlinearity condition starts to fail (since the maximum amplitude of the excitation increases substantially)
making more prominent the existence of higher harmonics and leading to effects such as an asymmetric time propagation.

 \begin{center}
 \textbf{V. BREATHER-LIKE EXCITATIONS} 
\end{center}

Having constructed an effective DNLS model we can finally pose the question of the existence of  breather-like solutions in our system in the regime of degeneracy.
If the coefficients $C,D$ of eq. (\ref{dnls1})  were zero, our system could be described within the so-called anticontinuum limit
of weakly coupled oscillators, since the linear coupling is already very small. In this limit, it has been  proven
that discrete breather solutions exist if there is a substantial degree of anharmonicity and no resonances
with the linear spectrum \cite{MacKay1994}. In our case with $C,D \neq 0$ one can still search for breathers provided, as well,
that their frequency does not belong to the phonon spectrum \cite{Oster2003}.
To this end the frequency in the ansatz of eq. (\ref{ans1}) should be changed from $\omega_0$ to $\omega<\omega_0$ out of the phonon band, i.e.
\begin{equation}
 x_n(t)=\Psi_n(t) e^{-i\omega t}+\Psi_n^*(t) e^{i\omega t}.
\end{equation}

\begin{figure}[htbp]
\begin{center}
\includegraphics[width=\columnwidth]{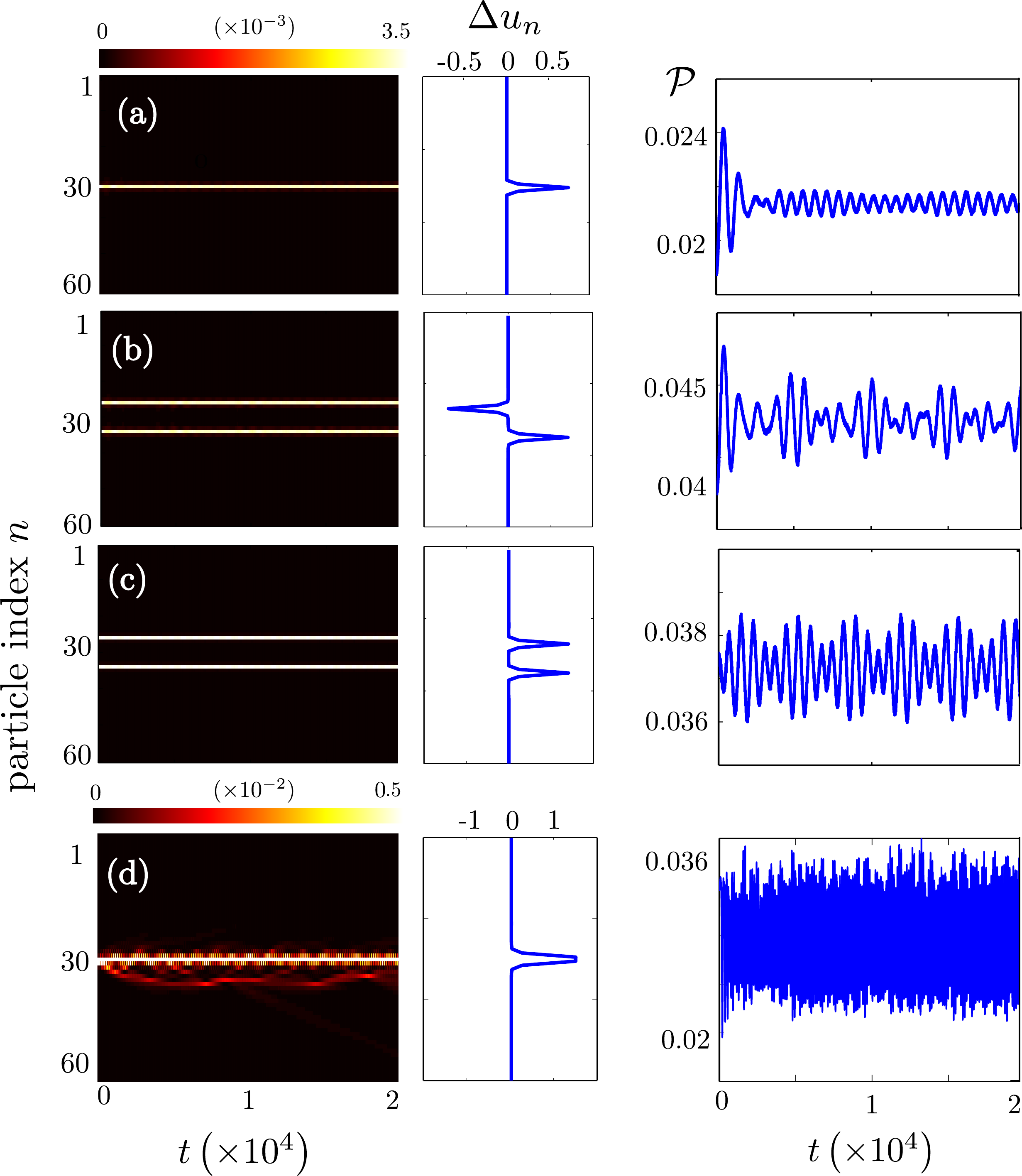}
\end{center}
\caption{\label{fsol1} (color online). (a)-(d) Time propagation of breather-like excitations  in the degeneracy regime,
at $r=r_3$ in (a), (b), (d) and  $r=r_4$ in (c).
The color plots of the first column depict the values of the local energy $E_n$ whereas the second column provides the initial profiles of the excitation
in  position coordinates, i.e. $\Delta u_n=u_n-u_n^{(0)}$ as a function of the particle index $n$. 
The third column illustrates the time evolution of the participation ratio $\mathcal{P}$ for the degenerate geometries and the initial conditions of (a)-(d). In  Figs.
(a)-(d) the initial excitation comes only from the displacement $\Delta u_n$ of particles from their equilibrium.  }
\end{figure}

 Repeating the steps of the previous section assuming that still
the RWA holds, we arrive at eq. (\ref{dnls2}) with an additional local term $\Lambda \Psi_n$
\begin{align}\label{dnls2}
&i\dot{\Psi}_n= A' (\Psi_{n-1}+\Psi_{n+1})+\Lambda \Psi_n +B'\abs{\Psi_n}^2\Psi_n \nonumber \\
& +C' \left[2\Psi_n\left(\abs{\Psi_{n-1}}^2+\abs{\Psi_{n+1}}^2\right)+\Psi_n^*(\Psi_{n-1}^2+\Psi_{n+1}^2)\right]\nonumber \\ 
&+ D' \left[2\abs{\Psi_n}^2\left(\Psi_{n-1}+\Psi_{n+1}\right)+\Psi_n^2\left(\Psi_{n-1}^*+\Psi_{n+1}^*\right)\right.\nonumber \\
&\left.+\abs{\Psi_{n-1}}^2\Psi_{n-1}+\abs{\Psi_{n+1}}^2\Psi_{n+1}\right] 
\end{align}
where $\Lambda=\left(\omega_0^2-\omega^2\right)/2\omega$, $A'=A\frac{\omega_0}{\omega}$, $B'=B\frac{\omega_0}{\omega}$, $C'=C\frac{\omega_0}{\omega}$ 
and $D'=D\frac{\omega_0}{\omega}$. In the anticontinuum limit $A',C',D'=0$ this equation can be easily checked to have the solutions 
$\Psi_n \in \{0, \sqrt{\frac{\Lambda}{B}} e^{i\phi_n} \}$ with arbitrary phases $\phi_n \in \mathbb{R}$.  
Simple real breather solutions can thus be constructed iteratively as strings of the elements $0$ and $\pm\sqrt{\frac{\Lambda}{B}}$.
Starting from such a breather solution and following the solution trajectories in the parameter space $A',C',D'$ with a Newton method
 \cite{Marin1996,Oster2003} we can find possible breather solutions for different values of $r$ corresponding to the geometries of interest. 
 The natural step then is to transform back to the $x_n,p_n$ coordinates and use these as an initial excitation in our system, simulating their time evolution
 within the full model equations as done in section III.

Our results are presented in Fig. \ref{fsol1} for different initial conditions at the degenerate geometries.  Evidently, many of them  (Figs. \ref{fsol1}(a)-(c))
keep their solitonic character in the presence of the full Coulomb interactions, a fact that  can be justified also by inspecting the time evolution of the participation ratio.
Indeed, the local energy profile of the excitations changes negligibly with oscillating potential and kinetic energy parts,
mapping to oscillating displacements $\Delta u_n$
and momenta $p_n$. 
These breather-like excitations include a single particle excitation (Fig. \ref{fsol1}(a)) as well as two-particle excitations with both opposite (Fig. \ref{fsol1}(b))
and equal displacements (Fig. \ref{fsol1}(c)).

Surprisingly enough, we have observed that the excited particles should always be separated by a distance (Figs. \ref{fsol1}(b),(c)) for the solitonic character to persist,
otherwise, if neighboring particles are excited there is always a small dispersion and a part of the excitation 
that focuses on a single particle, acquiring finally a rather asymmetric profile (Fig. \ref{fsol1}(d)). 
This picture is also valid within the DNLS framework where such solutions are found to undergo  a spontaneous symmetry breaking in the course of propagation.

 \begin{center}
 \textbf{VI. SUMMARY AND CONCLUSIONS}  
\end{center}

We have shown that a system of  charged particles confined on  a toroidal helix
can react in qualitatively different ways when exposed to an initial excitation, depending on the geometric properties 
of the confining manifold. In particular, while dispersion is the major feature of the dynamics for very small helix radii $r$,
the excitation  self-focuses and localizes for values of $r$ in the so-called degeneracy regime, where the linear coupling is very small.
Beyond this regime, the time evolution of the excitation is characterized by a defocusing which is gradually dominated again by dispersion.
Especially the self-focusing of the excitation observed for the degenerate geometries constitutes a hallmark of the existing nonlinearity in the system.
Interestingly enough, the nonlinear part of the interactions (both local and nonlocal) does not approach zero in this regime, nor does it undergo a change in sign with
increasing $r$,
contrary to the behaviour of the respective linear part. It is primarily this contrast between the robustness of the nonlinearity and the variability of the linear part 
which induces the variety of dynamics in the system, allowing for its control through the tuning of the underlying geometry.

Furthermore, we have  identified the character of the leading nonlinear terms
and constructed an effective discrete nonlinear Schr\"{o}dinger  model with additional nonlinear couplings 
which has allowed us to predict and  interpret the different responses of the helical chain to its excitation.
Through this model it has also been possible to identify some special breather-like excitations in our system which propagate in time keeping their  shape.
Overall, we emphasize that the present setup offers many possibilities for the implementation of various nonlinear models ranging from the Fermi-Pasta-Ulam  to different 
discrete nonlinear Schr\"{o}dinger models with both attractive and repulsive nonlinearities only by the tuning of a single geometrical parameter, i.e. the radius $r$ of the helix.

Regarding its possible experimental realization, it relies on the challenging task of constructing a helical trap for charged particles.
Although such a construction might not be straightforward, there have been certain advances towards this direction in different contexts, such as the realization
of free-standing helical nanostructures \cite{Prinz2000,Schmidt2001,Gogotsi2006} and the design of helical traps for neutral atoms \cite{Vetsch2010, Reitz2012}.
Moreover, our results could  be  of relevance in studies of energy and excitation transfer  in helical molecules, such as DNA. 
Finally, macroscopic realizations of our system could also be possible by using for example charged
beads as done for the study of polymers \cite{Reches2009,Tricard2012}.

 \begin{center}
 \textbf{APPENDIX: DEFINITION OF LOCAL ENERGIES}  
 \end{center}

 As briefly mentioned in Sec. III the definition of local energies $E_n$ in a system of 
 interacting particles is not unique.  For a general potential $V=\frac{1}{2}\sum_{i,j, i \neq j}^{N} W(u_i,u_j)$ the most common partition
 consists in equally distributing each part of the potential energy among the particles involved 
 \begin{equation}
  E_n=\frac{1}{2}\sum_{i, i\neq n}^{N} W(u_i,u_n)+K_n-E_n^{(0)},
 \end{equation}
 with $E_n^{(0)}=\frac{1}{2}\sum_{i, i\neq n}^{N} W(u_i^{(0)},u_n^{(0)})$
 the local energy of the equilibrium configuration $\{u_i^{(0)}\}$ and $K_n$ the kinetic energy of the particle $n$. Although such 
 a definition is very useful in describing systems with nearest neighbor interactions such as
 oscillator chains \cite{Sarmiento1999,Hennig2007}, it  can lead to  problems concerning the description of long-range interacting systems and particularly
 those involving Coulomb interactions, as the one we consider in this paper. The major problem is that it generally leads to
 both positive and negative values of $E_n$ and thus demands a handling of both positive and negative excitations.  Negative local energies 
 should be avoided if we wish to interpret the local energy as the amount or 
probability of the corresponding particle being excited and proceed in defining  quantities such as the participation ratio $\mathcal{P}$ of eq. (\ref{prat}).
 
 \begin{figure}[t!]
\begin{center}
\includegraphics[width=\columnwidth]{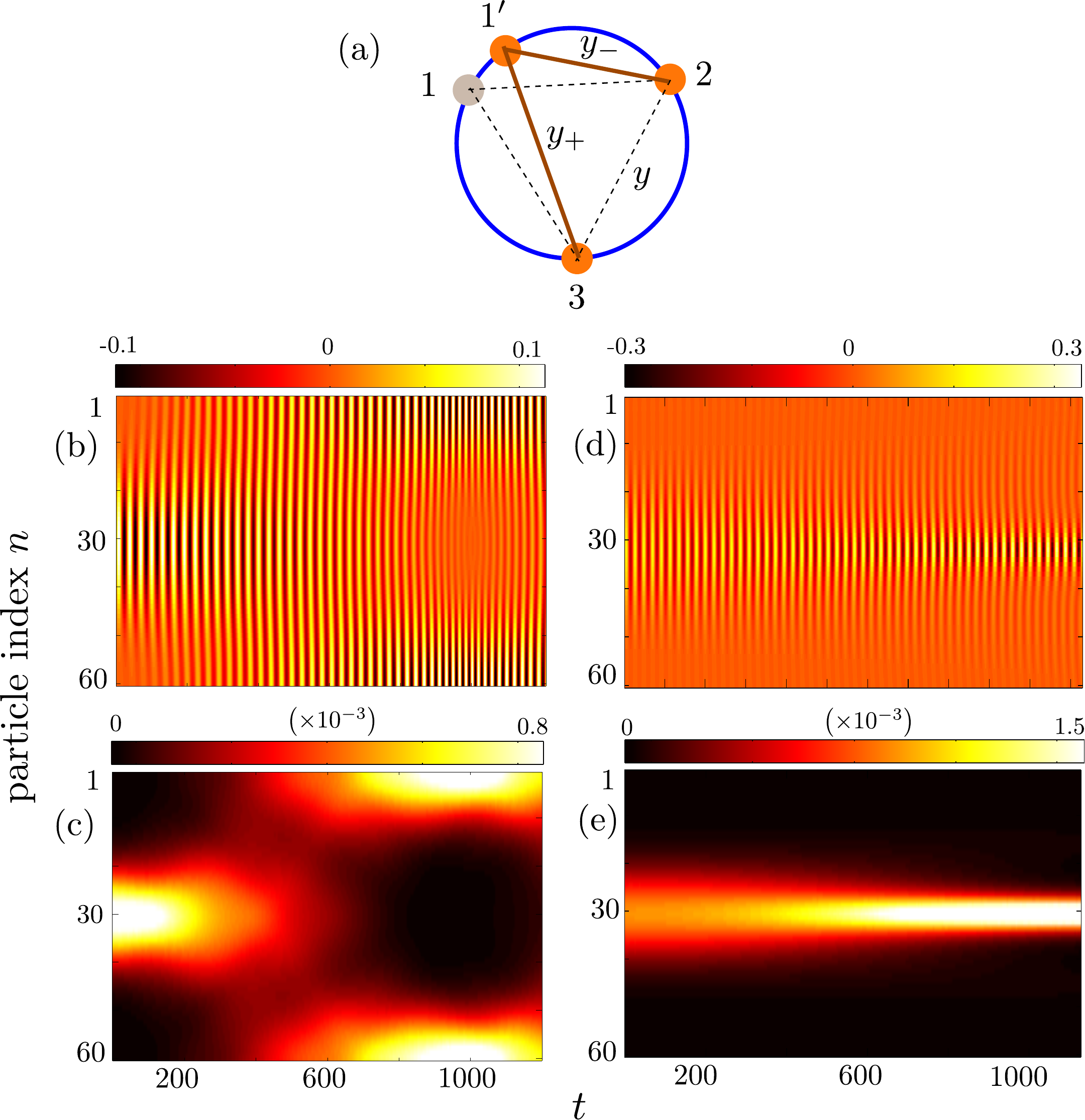}
\end{center}
\caption{\label{exa1} (color online). (a) A schematic illustration of three charges on a ring at equilibrium (1,2,3) and after 
particle 1 has been displaced ($1'$,2,3). The dashed lines denote the Euclidean distances in equilibrium, whereas the solid ones, the distances after the displacement.
(b),(c) The time evolution of  a Gaussian excitation for $r=r_1$. (d),(e) The time evolution of  a Gaussian excitation for $r=r_2$.
The colors in (b),(d) encode  the values of the displacement from equilibrium $x_n$ whereas in (c),(e) the values of the local energies $E_n$ defined in eq. (\ref{defle2}).
The figures (b)-(e) show simulation results for $N=60$ particles and $\nu=1/2$.}
\end{figure}
 
 Let us demonstrate this fact with a simple example of a Coulomb system consisting of three identical charges moving on a ring (Fig. \ref{exa1}(a)).
 The equilibrium configuration of such as system is the equidistant one with the charges sitting in the corners of an equilateral triangle
 with a side of length $y$. Evidently then $E_n^{(0)}=\frac{1}{y},~n=1,2,3$. If particle $1$ is clockwise displaced from equilibrium
up to  the point $1'$ the resulting local excitation energies would be
\[E_1=\frac{1}{2y_+}+\frac{1}{2y_-}-\frac{1}{y},~E_2=\frac{1}{2y_-}-\frac{1}{2y},~E_3=\frac{1}{2y_+}-\frac{1}{2y},\]
where $y_+>y>y_-$. Obviously $E_1,E_2>0$, whereas $E_3<0$, demonstrating that following this definition the local energy can become negative. 
Another minor disadvantage of this definition of local energy for long range interacting particles  is that it is highly non-local, meaning that if a single particle is displaced
from equilibrium, the local energies of other particles even those far apart will acquire a finite value.

A way to fix this last issue is to construct the local energy $E_n$ so that it contains: 
all the parts of the potential depending solely on $u_n$, none of the parts that are independent of $u_n$ and 
half of the terms involving both $u_n$ and any other $u_i$.
This can be written in a compact form
 \begin{align}\label{tayap2}
E_n&=\frac{1}{2}\sum_{i, i\neq n}^{N} [W(u_n,u_i)+W(u_n,u_i^{(0)})] \nonumber \\
&-\frac{1}{2}\sum_{i, i\neq n}^{N} [W(u_n^{(0)},u_i)+W(u_n^{(0)},u_i^{(0)})]+K_n,
\end{align}
which can be checked to sum up to the total excitation energy, i.e. $\sum_{n=1}^{N}E_n=E$.
Such a definition solves the problem of nonlocality since it distributes all single-particle potential terms 
to the corresponding particles, yielding for example $E_n=0$ for $u_n=u_n^{(0)}$ and $K_n=0$.
It can  also be checked to solve the problem of negativity in many cases, such as the one of the above three-particle example.
There are however still special cases in which negative local energies can be encountered,
 where the negativity originates  from the harmonic part of the total potential contained in each $E_n$
 \begin{align}
  V^{(2)}_n&=\frac{1}{2}\left.\frac{\partial^2 V}{\partial u_n^2 }\right|_{{(0)}}\left(u_n-u_n^{(0)}\right)^2 \nonumber \\
  &+\frac{1}{2}\sum_{i, i\neq n}^{N}\left.\frac{\partial^2 V}{\partial u_n \partial u_i }\right|_{{(0)}}\left(u_n-u_n^{(0)}\right)\left(u_i-u_i^{(0)}\right) 
 \end{align}
which can assume negative values at a given $n$.
To fix this, one can make use of the overall stability of the equilibrium which
leads to the existence of a different decomposition of the harmonic contributions to the potential into purely positive parts
in the form \cite{Allen1998}
\begin{equation}
 \tilde{V}^{(2)}_n=\frac{1}{2}\left(\sum_{j=1}^{N}\Omega_{nj} \left(u_j-u_j^{(0)}\right)\right)^2,
\end{equation}
where $\Omega$ denotes the square root of the Hessian. 
We thus adjust our definition (\ref{tayap2}) by subtracting $V_n^{(2)}$ and adding instead $\tilde{V}^{(2)}_n$, i.e.
\begin{align}\label{defle2}
E_n&=\frac{1}{2}\sum_{i, i\neq n}^{N} [W(u_n,u_i)+W(u_n,u_i^{(0)})] \nonumber \\
&-\frac{1}{2}\sum_{i, i\neq n}^{N} [W(u_n^{(0)},u_i)+W(u_n^{(0)},u_i^{(0)})]  \nonumber \\
&+K_n-V^{(2)}_n+ \tilde{V}^{(2)}_n.
\end{align}

This definition provides a sufficiently local (although not entirely, due to the term  $\tilde{V}^{(2)}_n$),
positive decomposition of the excitation energy which still satisfies $\sum_{n=1}^{N}E_n=E$.
Its use can be further justified by comparing the time evolution of $E_n$ 
with the time evolution of the displacement of the particle $n$ from equilibrium $x_n=s_n-s_n^{(0)}$. A comparison of Figs. \ref{exa1}(b),(d) with \ref{exa1}(c),(e) respectively,
leads  to the conclusion that the time evolution of $E_n$ captures nicely all the features of the propagating excitation such as the focusing and the spreading 
and only filters out the fast oscillations in Figs. \ref{exa1}(b),(d) caused by the fast continuous conversion of potential to kinetic energy and vice versa. 
\begin{center}
{ \textbf{ACKNOWLEDGEMENTS}}
\end{center}
A. Z. thanks the International Max Planck Research School for Ultrafast Imaging and Structural Dynamics for a PhD scholarship.
J. S. gratefully acknowledges support from the Studienstiftung des deutschen Volkes.


\begin{thebibliography}{99}



\bibitem{Scott1999} A. C. Scott, \textit{Nonlinear Science}, Oxford University Press, Oxford,  (1999).

\bibitem{Akhmanov1968} S. A. Akhmanov, A. P. Aukhorukov and R. V. Khokhlov, Sov. Phys. Usp. \textbf{10}, 609 (1968).
\bibitem{Eilbeck1985} J. S. Eilbeck, P.S. Lombdahl and A. C. Scott, Physica D \textbf{16}, 318 (1985).
\bibitem{Johansson1995} M. Johansson, M. H\"{o}rnquist and R. Riklund, Phys. Rev. B \textbf{52}, 231 (1995).

\bibitem{Sievers1988} A. J. Sievers and S. Takeno, Phys. Rev. Lett. \textbf{61}, 970 (1988).
\bibitem{Remoissenet1999} M. Remoissenet, \textit{Waves Called Solitons: Concepts and Experiments}, Springer-Verlag, Berlin,  (1999).
\bibitem{Campbell2004} D. K. Campbell, S. Flach and Yu S. Kivshar, Phys. Today \textbf{57}, 43 (2004).


\bibitem{Kevrekidis2009} P. G. Kevrekidis, \textit{The Discrete Nonlinear Schr\"{o}dinger Equation}, Springer-Verlag, Heidelberg, (2009).

\bibitem{Christodoulides1988} D. N. Christodoulides and R. I. Joseph, Opt. Lett. \textbf{13}, 794 (1988).
\bibitem{Morandotti1999} R. Morandotti \textit{et al.}, Phys. Rev. Lett. \textbf{83}, 4756 (1999).
\bibitem{Sukhorukov2003} A. A. Sukhorukov, Yu S. Kivshar, H. S. Eisenberg and Y. Silberger, IEEE \textbf{39}, 31 (2003).

\bibitem{Trombettoni1997} A. Trombettoni and A.Smerzi, Phys. Rev. Lett. \textbf{79}, 4950 (1997).
\bibitem{Abdullaev2001} F. Kh. Abdullaev \textit{et al.}, Phys. Rev. A \textbf{64}, 043606 (2001).
\bibitem{Hennig2010} H. Hennig, J. Dorignac and D. K. Campbell, Phys. Rev. A \textbf{82}, 053604 (2010).

\bibitem{Mingaleev1999} S. F. Mingaleev \textit{et al.}, J. Biol. Phys. \textbf{25}, 41 (1999).
\bibitem{Peyrard2004} M. Peyrard, Nonlinearity \textbf{17}, R1 (2004).
\bibitem{Koko2012} A. Dang Koko \textit{et al.}, Chaos \textbf{22}, 043110 (2012).

\bibitem{Gaididei2000} Yu B. Gaididei, S. F. Mingaleev and P. L. Christiansen, Phys. Rev. E \textbf{62}, 53(R) (2000).

\bibitem{Archilla2001} J. F. R. Archilla, P. L. Christiansen and Yu B. Gaididei, Phys. Rev. E \textbf{65}, 016609 (2001).
\bibitem{Sanchez2002} B. S\'anchez-Rey, J. F. R. Archilla,  F. Palmero and F. R. Romero, Phys. Rev. E \textbf{66}, 017601 (2002).



\bibitem{Kevrekidis2004} P. G. Kevrekidis  \textit{et al.}, Phys. Rev. E \textbf{70}, 066627 (2004).
\bibitem{Takeno2005} S. Takeno, S. V. Dimitriev, P. G. Kevrekidis and A. R. Bishop, Phys. Rev. B \textbf{71}, 014304 (2005).

\bibitem{Zampetaki2015} A. V. Zampetaki, J. Stockhofe and P. Schmelcher, Phys. Rev. A \textbf{91}, 023409 (2015).

\bibitem{Oster2003} M. \"{O}ster, M. Johansson and A. Eriksson, Phys. Rev. E \textbf{67}, 056606 (2003).
\bibitem{Oster2005} M. \"{O}ster and M. Johansson, Phys. Rev. E \textbf{71}, 025601(R) (2005).




\bibitem{Sarmiento1999} A. Sarmiento, R. Reigada, A. H. Romero and K. Lindenberg,  Phys. Rev E \textbf{60}, 5317 (1999).
 \bibitem{Hennig2007} D. Hennig, S. Fugmann, L. Shimansky-Geier and P. H\"{a}nggi,  Phys. Rev E \textbf{76}, 041110 (2007).

\bibitem{Allen1998} P. B. Allen and J. Kelner, Am. J. Phys. \textbf{66}, 497 (1998).

\bibitem{Zampetaki2013}  A. V. Zampetaki, J. Stockhofe, S. Kr\"{o}nke and P. Schmelcher, Phys. Rev. E \textbf{88}, 043202 (2013).

\bibitem{Fermi1995} E. Fermi, J. Pasta and S. Ulam, Los Alamos, Report No. LA-1940 (1955);
G. P. Berman and F. M. Izrailev, Chaos \textbf{15}, 015104 (2005).

\bibitem{Kivshar1992}  Yu. S. Kivshar and M. Peyrard, Phys. Rev. A \textbf{46}, 3198 (1992).

\bibitem{Bender1978}  C. M. Bender and S. A. Orszag, \textit{Advanced Mathematical Methods for Scientists and Engineers}, McGraw-Hill, Inc. (1978).

\bibitem{MacKay1994} R. S. MacKay and S. Aubry, Nonlinearity \textbf{7}, 1623 (1994).

\bibitem{Marin1996} J. L. Mar\'{i}n and S. Aubry, Nonlinearity \textbf{9}, 1501 (1996).

\bibitem{Prinz2000} V. Y. Prinz \textit{et al.}, Physica E \textbf{6}, 828 (2000).

\bibitem{Schmidt2001} O. G. Schmidt and K. Eberl, Nature \textbf{410}, 168 (2001).

\bibitem{Gogotsi2006} \textit{Nanotubes and Nanofibers} edited by Y. Gogotsi (Taylor \& Francis Group, LLC New York, 2006).

%
%
%
%
%
%

\bibitem{Vetsch2010} E. Vetsch \textit{et al.}, Phys. Rev. Lett. \textbf{104}, 203603 (2010).


\bibitem{Reitz2012} D. Reitz and A. Rauschenbeutel, Opt. Comm. \textbf{285}, 4705 (2012).

\bibitem{Reches2009} M. Reches, P. W. Snyder, G. M. Whitesides, Proc. Nat. Acad. Sci.  \textbf{106}, 17644 (2009).

\bibitem{Tricard2012} S. Tricard \textit{et al.}, Phys. Chem. Chem. Phys.  \textbf{14}, 9041 (2012).



\end{thebibliography}
\end{document}